\documentclass{iau}

\usepackage{amsmath}
\usepackage{graphicx}
\usepackage{multirow}
\usepackage{hyperref}

\begin{document}

\lefttitle{Masters \& Stark}
\righttitle{HI-MaNGA: Results from (21cm-HI) single-dish observations of MaNGA Survey Galaxies}

\jnlPage{1}{4}
\jnlDoiYr{2025}
\doival{10.1017/xxxxx}

\aopheadtitle{HI-MaNGA}
\editors{D.J. Pisano, Moses Mogotsi, Julia Healy, Sarah Blyth}

\title{HI-MaNGA: Results from (21cm-HI) single-dish observations of MaNGA Survey Galaxies}

\author{Karen Masters$^1$, David Stark$^2$}
\affiliation{1. Depts of Physics \&  Astronomy, Haverford College, 370 Lancaster Ave, Haverford, PA 19041, USA}
\affiliation{2. Space Telescope Science Institute, 3700 San Martin Dr Baltimore, MD, 21218, USA}

\begin{abstract}
In a poster presentation for IAU Symposium 392: ``Neutral hydrogen in and around galaxies in the SKA era", we gave an overview of the HI-MaNGA project which is working to obtain complementary information about the cold gas (neutral hydrogen traced by the radio 21cm line) content of Mapping Nearby Galaxies at Apache Point Observatory (MaNGA) sample galaxies. MaNGA, part of the fourth incarnation of the Sloan Digital Sky Surveys (SDSS-IV), obtained spatially resolved spectral maps for 10,000 nearby galaxies selected to create a representative sample out of the SDSS Main Galaxy Sample. MaNGA data have provided a census of the stellar and ionized gas content of these galaxies, as well as kinematics of both stars and gas. 
Adding HI information via the HI-MaNGA program, which has observed or collected 21cm line data for 70\% of the full MaNGA sample, has been crucial for a number of applications, but especially understanding the physical mechanisms that regulate gas accretion, and through that star formation and quenching of star formation.

\end{abstract}

%\begin{keywords}
%Key1, Key2, Key3, Key4
%\end{keywords}

\maketitle

\section{Introduction}

The Mapping Nearby Galaxies at Apache Point Observatory (MaNGA) project \citep{Bundy2015} which was part of the fourth incarnation of the Sloan Digital Sky Surveys \cite[SDSS-IV,][]{Blanton2017} used an optical integral field-unit (IFU) system \citep{Drory2015} on the Sloan Foundation Telescope from 2014-2020 to observe a sample of 10,010 unique galaxies selected from the SDSS Main Galaxy Sample \citep{Strauss2002}. These galaxies provide a representative sample of nearby galaxies ($z\sim0.03$ and $\log{M_\star/M_\odot}\sim 9-12$). For more details on the sample selection of MaNGA see \citet{Wake2017}, other technical details on MaNGA observations can be found in \citet{Law2015,Yan2016}.

MaNGA resolved optical spectroscopy provide a wealth of data on the stellar and ionized gas properties of nearby galaxies at a spatial sampling of roughly 1-2 kpc (2" at the typical distance of the galaxies). This information provides details accounting of the stellar masses, ages and metallicities, as well as motions of the stars and ionized gas, and information on ionization properties and more from the emission lines \citep{Law2016,Westfall2019,Belfiore2019}. However galaxies also contain cold gas \citep{Saintonge2022}, the largest component of which by mass is the neutral hydrogen (or HI), which can be observed via its 21cm (1.4 GHz) line emission using a radio telescope.

\section{HI-MaNGA Observations}
Since 2016, the HI-MaNGA project \citep{Masters2019} has made use of over 2500 hours of filler time\footnote{Under GBT project codes: 16A-095, 17A-012, 19A-127, 20B-033, 21B-130 and most recently approved as 24B-263} on the Robert C. Byrd Green Bank Telescope (GBT) to observe over 4000 galaxies in the MaNGA sample in the 21cm-line from HI (neutral hydrogen). HI-MaNGA also adds existing data for MaNGA galaxies from the ALFALFA (Arecibo Legacy Fast Arecibo L-band Feed Array) HI survey \citep{Haynes2018} and is working on including data from The FAST All Sky HI survey \citep[FASHI; ][]{Zhang2024} and the HI mass inventory for the REsolved Spectroscopy Of a Local VolumE (RESOLVE) survey \citep{Stark2016}. At the time of the 2024 IAU General Assembly, the HI-MaNGA project had observed or collected total HI masses and rotation widths (for detections), or HI mass upper limits (for non-detections), for over 70\% of the MaNGA sample. 

\begin{figure}[h]
  \centering
\includegraphics[scale=.3]{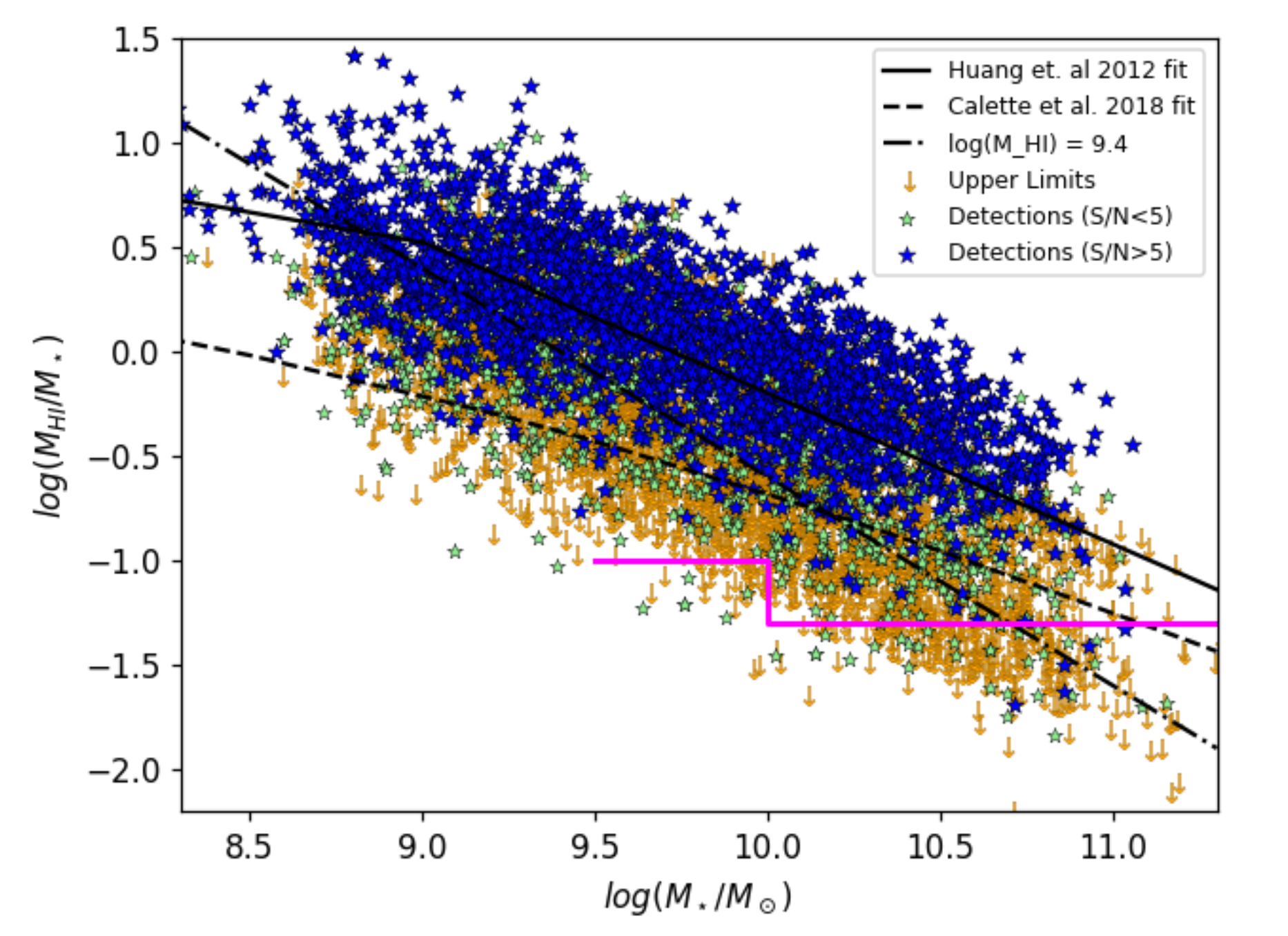}
  \caption{The HI Mass Fraction Plotted against stellar mass for all GBT and ALFALFA data in a preliminary version of the HI-MaNGA DR4 sample (over 7000 MaNGA galaxies). Diagonal ines are shown for a constant HI mass of $\log(M_{\rm HI}/M_\odot=9.4$, as well as published fits to the relation from ALFALFA \citep{Huang2012} and a compilation of HI-surveys \citep{Calette2018}. The horizontal (magenta) lines show targets for fixed upper limits in new observations. }
  \label{fig:hifractionmass}
\end{figure}

Our initial goal, as described in \cite{Masters2019}, was to obtain HI (21cm-line) followup for all galaxies in the MaNGA Survey main sample to $z<0.05$. To date the project has had three large data releases: DR1 \citep{Masters2019}; DR2 \citep{Stark2021} and DR3, which came out as part of the SDSS-DR17 \citep{DR17}. Figure \ref{fig:hifractionmass} show the HI mass fraction versus stellar mass \citep[from the Pipe3D analysis of MaNGA; ][]{Sanchez2022} for a preliminary version of the DR4 data release containing detections or upper limits for just over 7000 MaNGA galaxies, meeting the initial goal. The DR4 data release is planned to coincide with the next Sloan Digital Sky Survey release. 

The latest HI-MaNGA data can always be obtained from the HI-MaNGA website hosted at the GBT\footnote{\tt https://greenbankobservatory.org/science/gbt-surveys/hi-manga/} or the SDSS-IV Value-Added-Catalog page\footnote{\tt https://www.sdss4.org/dr17/manga/hi-manga}.

\section{Recent Publications Using HI-MaNGA data}
 The poster also provided information about recent publications using HI-MaNGA data, from members of the HI-MaNGA team, including: 
 \begin{itemize}
 \item \citet{Sharma2023}. ``HI-rich but low star formation galaxies in MaNGA: physical properties and comparison to control samples." In this paper the authors found that 5\% of HI-detections in MaNGA were in quenched galaxies (low specific star-formation rate (sSFR) as indicated by infra-red photometry from WISE). \citet{Sharma2023} presented evidence that this was caused by mixture of recent HI accretion (identified via measures of increased kinematic misalignment), active galactic nuclei (AGN) feedback and/or secular evolution from bars, or that the HI was perhaps found in high angular momentum (low density) discs. This suggested that resolved HI follow-up would be of interest. 
 \item \citet{Salem2024}, ``Finding Passive Galaxies in HI-MaNGA: The Impact of Star Formation Rate Indicator” - in this short followup paper to \citet{Sharma2023} the authors  demonstrate similar results using a variety of different sSFR measures. 
     \item \citet{Goddy2023}: ``A comparison of the baryonic Tully-Fisher relation in MaNGA and IllustrisTNG." In this paper, HI-MaNGA data were used in combination with MaNGA data to compare baryonic Tully-Fisher relations to those of simulated galaxies
     \item \citet{Frank2023}: ``The HI content of red geyser galaxies." This paper used HI-MaNGA observations of MaNGA red geysers \citep{Cheung2016} and stacked them to measure typical HI content of mostly undetected red geyser galaxies. 
    \item \citet{Shapiro2022}:  ``Testing Algorithms for Identifying Source Confusion in HI-MaNGA.". This paper, done in collaboration with the Apertif  team was a test of how source algorithms used first in HI-MaNGA by \citet{Stark2021} perform compared to HI data which more angular resolution from Apertif on the Westerbork Synthesis Radio Telescope \citep[WSRT,][]{Apertif}.
    \item \citet{Stark2021}: ``HI-MaNGA: Tracing the physics of the neutral and ionized ISM with the second data release". This paper, as well as being the DR2 release for HI-MaNGA, used the data to examine relationships between HI-to-stellar mass ratio ($M_{\rm HI}/M_\star$) and average ISM/star formation properties probed by optical emission lines. 
    \item \citet{Goddy2020}: ``L-band Calibration of the Green Bank Telescope from 2016--2019”. This paper used calibration data taken in the first four years of HI-MaNGA observations to reveal a constant offset of 20\% in the calibration relative to the standard values given in the GBT data reduction pipeline. 
 \end{itemize}

\section{Conclusion and Plans for the Future}
HI-MaNGA is meeting its goal of obtaining single dish HI data to complement the MaNGA survey \citep{Bundy2015, Masters2019}. In addition to the papers noted here, HI-MaNGA data release papers \citep{Masters2019, Stark2021} have almost 100 citations demonstrating the use of this data to the community. 

The next data release from HI-MaNGA (DR4), will contain over 7000 HI observations (detections and upper limits) for MaNGA galaxies, and is planned to coincide with the next Sloan Digital Sky Survey release (as a Value Added Catalogue). The latest data can always be obtained from the HI-MaNGA website hosted at the GBT\footnote{\tt https://greenbankobservatory.org/science/gbt-surveys/hi-manga/} or the SDSS-IV Value-Added-Catalog page\footnote{\tt https://www.sdss4.org/dr17/manga/hi-manga}. 

Meanwhile the project continues observing, with approved filler time under {\tt AGBT24A-263}, both to finish the initial survey, to re-observe targets with significant data loss from non-astronomical emissions (usually the L3 signal from GPS satellites) and improve our upper limits for non-detections. For these new observations, we add a new goal to observe to a fixed gas fraction limits for a subset of HI-MaNGA non-detections: aiming for $M_{\rm HI}/M_\star=0.05$, 0.1 for $\log(M_\star/M_\odot>10$, 9.5 respectively, for those targets where that will be possible in $t<3$hrs on source (see illustration of these limits in Figure \ref{fig:hifractionmass}).

The HI-MaNGA project provides an example of the kind of multi-wavelength collaboration that will be necessary to fully exploit the rich information we will have on the HI-content of nearby galaxies in the era of SKA observations of the the HI skies. Additionally, the HI-MaNGA project has handled a large amount of GBT filler time with a very small core team by working with GBT observers to run observations from scripts, and making simple data reduction wrappers enabling undergraduate researchers to contribute to data reduction.

\end{document}